\documentclass[12pt,eqsecnum]{revtex4}
\begin{document}

\title{ New forms of BRST symmetry in rigid rotor
  }


\author{ Sumit Kumar Rai \footnote{e-mail address: sumitssc@gmail.com}}
\author{Bhabani Prasad Mandal \footnote{e-mail address:
\ \ bhabani.mandal@gmail.com,  \ \ bhabani@bhu.ac.in \   }}


\affiliation{ Department of Physics,\\
Banaras Hindu University,\\
Varanasi-221005, INDIA. \\
}

\begin{abstract}
We derive the different forms of BRST symmetry by using the 
Batalin-Fradkin-Vilkovisky formalism in a rigid rotor. The so called ``dual-BRST" 
symmetry is obtained from usual BRST symmetry by making a canonical 
transformation in the ghost sector. On the other hand, a canonical transformation in the sector 
involving Lagrange multiplier and its corresponding momentum leads to a new form of 
BRST as well as dual-BRST symmetry.
\end{abstract}
\maketitle
\section{Introduction}
Becchi-Rouet-Stora and Tyutin (BRST) symmetry \cite{brst} provides a basis for the modern quantization of 
gauge theory and is very important tool in characterizing various 
renormalizable field theoretic models. Path integral quantization of gauge theories 
\cite{hete} can be 
done both in Lagrangian formulation and Hamiltonian formulation. In both the formulation, 
phase space is extended by incorporating the Grassmanian odd ghost variables where the gauge 
invariance is ensured by the BRST symmetry.

The Hamiltonian approach developed by Batalin, Fradkin and Vilkovisky (BFV) \cite {frvi} is a powerful 
technique to study the BRST quantization of constrained systems. The main features of BFV 
approach are: it does not require closure off-shell of the gauge algebra and therefore does 
not need an auxiliary field, this formalism heavily relies on BRST transformations which are 
independent of the gauge condition, this method is even applicable to Lagrangians which are 
not quadratic in velocities and hence is more general than the strict Lagrangian approach. 
Being based on the Hamiltonian, the approach is closer to Hilbert space techniques and to unitarity. This method uses an extended phase space where the Lagrange multiplier 
and the ghosts are treated as dynamical variables. The generator of BRST symmetry  for systems 
with first class constraints can be constructed from the constraints in a gauge independent 
way whose cohomology produces the physical states. 

A great deal of work has been done on various models using BFV approach for the systems with 
first class constraints such as QED, U(1) gauge theory etc. \cite{gara,nega,gaete}. This 
approach has also been 
applied to the systems with second class constraints such as Proca model, chiral Schwinger 
model etc \cite{proca,chsh} by converting them to first class 
constraints using various techniques. A great deal of work has also been done on the various form of the BRST symmetry such as the non-local and non-covariant symmetry for QED \cite{lamc}, the non-local but covariant symmetry for QED \cite{tafi} and the local but non-covariant symmetry in Abelian gauge theories \cite{yale}. In all the above mentioned symmetries, the variaton of gauge fixing part vanishes which is defined as ``dual'' to the vanishing of the variation of kinetic part in the usual BRST \cite{lamc}.  The so called dual-BRST( also called Co-BRST symmetry) where the 
gauge fixing part is independently invariant and the variation of the kinetic 
part cancels out with that of the ghost part of the effective action, was thought to be an 
independent symmetry. It has 
been shown  recently, using BFV formalism that the dual BRST symmetry is not an 
independent symmetry but 
can be obtained by canonical transformation in the ghost sector of BRST symmetry for U(1) gauge theory 
 \cite{gaete,yale}.

In this paper, our purpose is to render the content of BFV-BRST technique more easily 
accessible to the non-expert by establishing the connection between the constrained systems, 
 BRST and dual-BRST symmetries. In particular, we obtain the different forms of BRST symmetry 
by using BFV-BRST formulation in a simple system like rigid rotor. The canonical 
transformation in the ghost sector of the gauge fixed 
action gives rise to dual-BRST symmetry. On the other hand, a canonical transformation in the 
sector involving the Lagrange multiplier and its corresponding momentum leads to a new form 
of BRST/dual-BRST symmetries. Our goal is to put
this powerful technique of BFV approach into the framework of elementary 
quantum mechanics.

The plan of our paper is as follows. In Sec. II, we give a brief introduction to BFV formalism. Using this BFV approach, we generate the BRST symmetries in Sec. III-A. We obtain the new form of BRST symmetry in Sec. III-B. In Sec. IV, we have shown that the dual-BRST symmetry for a rigid rotor is obtained by making a canonical transformation in the ghost sector. A new form of dual-BRST symmetry is obtained in Sec. IV-A. Sec. V is devoted to conclusion and discussion.

\section{BFV formalism}
This method provides a general procedure to quantize systems with first class constraints. We 
recapitulate the essence of this approach in terms of finite number of phase space variables. 
The action under such considerations can be written as 
\begin{equation}
S=\int dt \left( p^\mu\dot{q}_\mu - H_0 -\lambda^a \Omega_a \right),
\end{equation}
where ($q^\mu,p_\mu$) are the canonical variables describing the theory. $H_0$ is the 
 Hamiltonian and $\lambda^a$ are the Lagrange multiplier associated with first class 
constraints, $\Omega_a$. In this approach, Lagrange multipliers $\lambda^a$ are dynamical 
variables and therefore, treated as the canonical variables. We introduce conjugate 
canonical momenta $p_\lambda^a$ to $\lambda^a$  where $p_\lambda^a$ must be imposed as new constraints 
such that the dynamics of the theory does not change. BFV method introduces a pair of 
canonically conjugate ghosts ( ${\cal{C}}^a,{\cal{P}}_a$) for each constraints of the theory. 
These ghosts follow the anticommutation relation as follows
\begin{equation}
\left \{{\cal{C}}^a(x,t) ,{\cal{P}}^b (y,t)\right\} = -i\delta^{ab}\delta ({\bf x}-{\bf y}),
\end{equation}
where ${\cal{C}}$ and ${\cal{P}}$  have ghost number 1 and -1 respectively.  The nilpotent generator, Q of BRST symmetry in an extended phase space of the system with 
first class constraints has the general form
\begin{equation}
Q={\cal{C}}_a \Omega^a + \frac{1}{2}{\cal{P}}^a f^{bc}_a {\cal{C}}_b {\cal{C}}_c, 
 \label{sg}
\end{equation}
where the $f^{bc}_a$ is a structure constant, $\Omega^a $ is the first class constraint.   
 According to the Fradkin-Vilkovisky theorem \cite{frvi}
which states that the generating functional in the extended phase space is given as 
\begin{equation}
Z_\Psi =\int {\cal{D}}\varphi \; \exp (iS_{eff}) \label{zpsi},
\end{equation}
where the effective action, $S_{eff}$ is 
\begin{equation}
S_{eff}=\int dt\left (p^\mu {\dot q}_\mu + {\dot{\cal{C}}}^a {\cal{P}}_a + p^a\dot{\lambda}_a 
-H_\Psi \right ). \label{seff}
\end{equation}
${\cal{D}}\varphi$ is the Liouville measure on the phase space. $H_\Psi$ is the extended 
Hamiltonian given as
\begin{equation}
H_\Psi =H_0 + \left\{Q,\Psi\right\}.
\end{equation}
$\Psi$ is the gauge fixed fermion and $Z_\Psi$ does not depend upon the choice of $\Psi$.
\section{BFV-BRST approach in rigid rotor}
\subsection{BRST symmetry}
We consider a rigid rotor in 2+1 dimension. The constraint equation is ($r-a)$=0. The canonical 
Hamiltonian for such a system can be written as \cite{brstp}
\begin{equation}
H_c =H_0 +\lambda (r-a),
\end{equation}
where $H_0=p^2_\theta/2mr^2$, and $p_\theta = (x p_y - y p_x)$. The action in a finite phase space can be written as
\begin{equation}
S=\int dt \left[ p_r \dot{r} +p_\theta \dot{\theta} - \frac{p_\theta ^2}{2mr^2} -\lambda (r-a)
\right ].
\end{equation}
Using Dirac's prescriptions \cite{dirac} for constraint analysis, it is trivial to see that 
the system has only two first class constraints, namely the primary constraint $p_\lambda$=0
  and the secondary constraint $(r-a)$=0. 

Using BFV approach, we introduce a pair of canonically conjugate ghosts (${\cal{C}}$,
${\cal{P}}$) with ghost number 1 and -1 respectively, for  the first class constraint, $p_\lambda$=0 
and another pair of canonically conjugate anticommuting ghosts ($\bar{{\cal{C}}},\bar{{\cal{P}}}$) with ghost 
number -1 and 1 respectively for other constraint, $(r-a)=0$. The effective action in the extended 
phase space using Eq. (\ref{seff}) becomes
\begin{equation}
S_{eff} = \int dt \left [ p_r {\dot r} + p_\theta {\dot{\theta}} + p_\lambda {\dot{\lambda
}} + {\dot{\cal{C}}} {\cal{P}}+ {\dot{\bar{\cal{C}}}} {\bar{\cal{P}}}- \frac {p_\theta 
^2}{2mr^2} -\left \{ Q,\Psi \right \}\right ]. \label{seff1}
\end{equation}
The symmetry generator for the rigid rotor from Eq. (\ref {sg}) is
\begin{equation}
Q_b=i\left[{\cal{C}}(r-a)+ {\bar{\cal{P}}}p_\lambda \right ]. \label{bc}
\end{equation}
Using the relation $\delta _b\phi={\left[\phi,Q_b\right]}_\pm $ ( + sign for bosonic and - for fermionic nature of $\phi$), the BRST charge given in Eq. 
(\ref{bc}) will generate the following BRST transformations
\begin{eqnarray}
 &&\delta _b p_r= {\cal{C}},\quad \quad \delta _b\lambda =-{\bar{\cal{P}}},\quad\quad \delta _b
{\bar{\cal{C}}}=p_\lambda , \nonumber \\
&&\delta _b \theta =0,\quad \quad  \ \delta _bp_\lambda =0,\quad \quad \;\;\; \delta _b p_\theta = 0,
\nonumber \\
&& \delta_b r =0, \quad \quad \ \delta _b {\cal{C}}=0, \quad \quad \;\;\;\; \delta _b {\bar{\cal
{P}}}=0,\nonumber\\
&&\delta _b {\cal{P}}=(r-a).\label{ghbrst}
\end{eqnarray}
We choose the gauge fixed fermion as 
\begin{equation}
\Psi= \left[{\cal{P}}\lambda +{\bar{\cal{C}}} \left (p_r +\frac{\xi}{2}p_\lambda\right )
\right ],
\end{equation}
and calculate
\begin{equation}
\left\{ Q_b, \Psi \right\} = \lambda (r-a) + {\bar{\cal{P}}}{\cal{P}}-{\cal{C}}{\bar{\cal
{C}}} + p_\lambda \left ( p_r+\frac{\xi}{2}p_\lambda\right ).
\end{equation}
Putting it into Eq. (\ref{seff1}), the effective action becomes
\begin{eqnarray}
S_{eff} &=&\int dt\left [ p_r {\dot r} + p_\theta {\dot{\theta}} + p_\lambda {\dot{\lambda
}} + {\dot{\cal{C}}} {\cal{P}}+ {\dot{\bar{\cal{C}}}} {\bar{\cal{P}}}- \frac {p_\theta 
^2}{2mr^2} -\lambda (r-a)\right. \nonumber \\
 &-&\left. {\bar{\cal{P}}}{\cal{P}}+{\cal{C}}{\bar{\cal{C}}} -p_\lambda \left ( p_r+\frac
{\xi}{2}p_\lambda \right )\right ]. \label{fseff}
\end{eqnarray}
The generating functional $Z_\Psi$ corresponding to the above effective theory can be written  as
\begin{eqnarray}
Z_\Psi &=&\int {\cal{D}}\varphi \exp \left[i\int dt \left\{ p_r {\dot r} + p_\theta {\dot{\theta}} + p_\lambda {\dot{\lambda
}} + {\dot{\cal{C}}} {\cal{P}}+ {\dot{\bar{\cal{C}}}} {\bar{\cal{P}}}- \frac {p_\theta 
^2}{2mr^2} -\lambda (r-a) \right. \right.\nonumber \\
&-& {\bar{\cal{P}}}{\cal{P}}+\left. \left. {\cal{C}}{\bar{\cal{C}}} -p_\lambda \left ( p_r+\frac
{\xi}{2}p_\lambda \right )\right \} \right].\label{zpsi1}
\end{eqnarray}
The effective action in Eq. (\ref{fseff}) is invariant under the BRST transformations given in Eq. (\ref{ghbrst}).
We integrate $Z_\Psi$ in Eq. (\ref{zpsi1}) over ${\cal{P}}$, ${\bar{\cal{P}}}$  to obtain
\begin{eqnarray}
Z_\Psi &=&\int {\cal{D}}\varphi^\prime \exp \left[i\int dt\left\{ p_r {\dot r} + p_\theta {\dot{\theta}}-\frac {p_\theta 
^2}{2mr^2} -\lambda (r-a)+p_\lambda (\dot{\lambda}-p_r) \right. \right. \nonumber\\
&-&\frac{\xi}{2}{p_\lambda}^2+\left. \left. {\cal{C}}{\bar{\cal{C}}}+{\dot{
\bar{\cal{C}}}} {\dot{\cal{C}}}\right\}\right ] \label{srigid}
\end{eqnarray}
and integrate Eq. (\ref{srigid}) over $p_\lambda$ to obtain
\begin{eqnarray}
Z_\Psi &=&\int {\cal{D}}\varphi^{\prime\prime} \exp \left[ i \int dt\left\{ p_r {\dot r} + p_\theta {\dot{\theta}}-\frac {p_\theta 
^2}{2mr^2} -\lambda (r-a)-\frac{1}{2\xi}{(\dot{\lambda}-p_r)}^2\right.\right.\nonumber \\
&+&\left.\left.{\cal{C}}{\bar{\cal{C}}}+{\dot{
\bar{\cal{C}}}} {\dot{\cal{C}}}\right\}\right ]. \label{srigid1}
\end{eqnarray}
which is same as the action mentioned in Ref. \cite{brstp}. The difference is that this approach does 
not require any auxiliary field and is done in an gauge independent way. The BRST symmetry 
after integrating over ${\cal{P}}$ and ${\bar{\cal{P}}}$ becomes 
\begin{eqnarray}
&&\delta _b p_r= {\cal{C}},\quad \quad \delta _b\lambda =-\dot{{\cal{C}}},\quad\quad \delta _b
{\bar{\cal{C}}}=p_\lambda , \nonumber \\
&&\delta _b \theta =0,\quad \quad \; \delta _bp_\lambda =0,\quad \quad \;\ \delta _b p_\theta = 0,\nonumber\\
&&\delta_b r =0,\quad \quad \; \delta _b {\cal{C}}=0,
\end{eqnarray}
which is similar to the BRST symmetries mentioned in Ref. \cite{brstp}.
 In the effective action given by Eq. (\ref{srigid1}), we observe that the term 
$\frac{1}{2\xi}{(\dot{\lambda} -p_r)}^2$ is like a gauge fixing term.
\subsection{New form of BRST symmetry}
In this section, we make a canonical transformation in $(p_\lambda,\lambda)$ sector as follows
\begin{eqnarray}
p_\lambda^\prime &=&p_\lambda+\frac{2}{\xi}p_r, \nonumber\\
\lambda^\prime &=&\lambda, \label{gtr}
\end{eqnarray}
to find a new form of BRST transformations.
  The path integral measure does not change as the Jacobian equals to 1 in this sort of transformation. The BRST transformation for the new variable $p_\lambda^\prime$ now becomes
\begin{equation}
\delta p_\lambda^\prime =\frac{2}{\xi}\delta p_r =\frac{2}{\xi} {\cal{C}}.
\end{equation}
The BRST transformation for $\bar{{\cal{C}}}$ is expressed in terms of new defined variable $p_\lambda^\prime$. This gives rise to new form of BRST symmetry \cite{rive}
\begin{eqnarray}
&&\delta _b p_r= {\cal{C}},\quad \quad \delta _b\lambda =-\dot{{\cal{C}}},\quad\quad  \delta _b \theta =0, \nonumber \\
&& \delta _b p_\theta = 0,\quad \quad \delta 
_b r =0,\quad \quad \quad \delta _b {\cal{C}}=0,\nonumber\\
&&\delta _b
{\bar{\cal{C}}}=p_\lambda +\frac{2}{\xi}p_r,\quad\quad \delta _bp_\lambda =-\frac{2}{\xi}\delta_b p_r =-\frac{2}{\xi} {\cal{C}},
\end{eqnarray}
which leaves the action given in Eq. (\ref{srigid}) invariant. It can be easily seen that they are nilpotent and can reduce to the original form of BRST symmetry.
\section{Dual-BRST symmetry}
We make the following canonical transformation in the ghost sector $ ({\cal{C}}^a,{\cal{P}}_a )$ of the theory
\begin{eqnarray}
{\cal{C}}\quad &\rightarrow &\quad {\cal{P}} \nonumber\\
{\cal{P}}\quad &\rightarrow &\quad {\cal{C}} \nonumber\\
{\bar{\cal{C}}}\quad &\rightarrow &\quad {\bar{\cal{P}}} \nonumber\\
{\bar{\cal{P}}}\quad &\rightarrow &\quad {\bar{\cal{C}}}
\end{eqnarray}
which does not change the effective action in Eq.  (\ref{fseff}). The generator of the BRST symmetry in Eq. (\ref{bc}) after the above canonical transformation becomes
\begin{equation}
Q_d= i \left[ {\cal{P}}(r-a)+{\bar{\cal{C}}}p_\lambda \right ].
\end{equation}
Now $Q_d$ generates the following new transformations
\begin{eqnarray}
 &&\delta _d p_r= {\cal{P}},\quad \quad \delta _d\lambda =-{\bar{\cal{C}}},\quad\quad \delta _d \theta =0,\nonumber \\
&& \delta _d p_\lambda =0,\quad \quad \delta _d p_\theta = 0,\quad \quad \;\; \; \delta 
_d r =0,\nonumber \\
&&\delta _d
{\bar{\cal{C}}}= 0 ,\quad\quad \; \delta _d {\cal{P}}=0, \quad\quad \quad \delta _d {\bar{
\cal{P}}}=p_\lambda, \nonumber\\
&&\delta_d {\cal{C}}=(r-a). \label{qdbrst}
\end{eqnarray}
The Jacobian of the canonical transformation in Eq. (\ref{qdbrst}) is unity, so
 the Liouville measure in the generating functional does not change. The effective action given in Eq. (\ref{srigid}) is symmetric under the transformations  mentioned in Eq. (\ref{qdbrst}).

We observe that the variation of gauge fixing part in Eq. (\ref{srigid1})  vanishes independently[i.e 
$\delta_d \left(\dot{\lambda}-p_r\right)=0 $ ]. This 
 is the dual-BRST symmetry as mentioned in the introduction part.  We carry out the integration over ${\cal{P}}$ and 
${\bar{\cal{P}}},$ in the generating functional given by Eq. (\ref{zpsi1}) to get the following dual-BRST symmetries
\begin{eqnarray}
&&\delta _d p_r= -{\dot{\bar{\cal{C}}}},\quad \quad \delta _d\lambda =-\bar{{\cal{C}}},\quad\quad 
\delta _d{\cal{C}}=(r-a) , \nonumber \\
&&\delta _d \theta =0,\quad \quad \quad \delta _d p_\lambda =0,\quad \quad\;\; \delta _d p_\theta = 0,\nonumber\\
&&\delta_d r =0,\quad \quad \quad \delta _d {\bar{\cal{C}}}=0,\label{dbrst}
\end{eqnarray}
under which the effective action given by Eq. (\ref{srigid1}) is invariant.
\subsection{New form of Dual-BRST symmetry}
The above mentioned dual-BRST symmetry is obtained by the canonical transformation in the ghost sector. A new form of  Dual-BRST symmetry can also be obtained by making a canonical transformation [Eq. (\ref{gtr})] in the sector of Lagrange multiplier and its momenta. Following the steps analogous to Sec III-B, we obtain the new form of dual-BRST symmetry in case of rigid rotor as 
\begin{eqnarray}
&&\delta _d p_r= -{\dot{\bar{\cal{C}}}},\quad \quad \delta _d\lambda =-\bar{{\cal{C}}},\quad\quad \delta _d \theta =0,
 \nonumber \\
&& \delta _d p_\theta = 0,\quad \quad \;\;\; \delta 
_d r =0,\quad \;\;\; \quad \delta _d {\bar{\cal{C}}}=0,\nonumber\\
&&\delta _d{\cal{C}}=(r-a), \;\;\delta _d p_\lambda =\frac{2}{\xi}\delta_d p_r =-\frac{2}{\xi}{\dot{\bar{\cal{C}}}}.\label{mdbrst}
\end{eqnarray}
Now $p_\lambda$ is changing non-trivially in the above transformations. The transformations mentioned in Eq. (\ref{mdbrst}) leave the action given in Eq. (\ref{srigid}) invariant.

\section{Conclusion}
We study the BFV-BRST formulation in a very simple system, rigid rotor to demonstrate the techniques to obtain different forms of BRST symmetries. Dual-BRST symmetry is obtained by making a canonical transformation in the ghost sector of the effective action of rigid rotor. This is the local and covariant version of the kind of transformations considered by Lavelle and McMullan \cite{lamc}, Tang and Finkelstein \cite{tafi} and Yang and Lee \cite{yale}. This implies BRST and dual-BRST are not independent symmetries in this case rather these are related through canonical transformation. On the other hand, a canonical transformation in the sector involving Lagrange multiplier and its momenta leads to a new form of BRST as well  as dual-BRST symmetry. This simple technique can be applied to more complicated system to derive different form of BRST as well as dual-BRST symmetry which can simplify the renormalizable program.

\vspace{.2in}

{\bf {\large Acknowledgment}}\\
\\We thankfully acknowledge the financial support from the Department of Science and Technology (DST), Government of India, under the SERC project sanction grant No. SR/S2/HEP-29/2007.

\end{document}